\documentclass[a4paper]{jpconf}

\usepackage{hyperref,amsfonts,amsmath,amssymb,graphicx,bm,subfigure}

\begin{document}
\title{Neutrino oscillations in gravitational waves}

\author{Maxim Dvornikov}

\address{Pushkov Institute of Terrestrial Magnetism, Ionosphere
and Radiowave Propagation (IZMIRAN),
108840 Moscow, Troitsk, Russia}

\ead{maxdvo@izmiran.ru}

\begin{abstract}
We study spin and flavor oscillations of neutrinos under the influence of gravitational waves (GWs). We rederive the quasiclassical equation for the evolution of the neutrino spin in various external fields in curved spacetime starting from the Dirac equation for a massive neutrino. Then, we consider neutrino spin oscillations in nonmoving and unpolarized matter, a transverse magnetic field, and a plane GW. We show that a parametric resonance can take place in this system. We also study neutrino flavor oscillations in GWs. The equation for the density matrix of flavor neutrinos is solved when we discuss the neutrino interaction with stochastic GWs emitted by coalescing supermassive black holes. We find the fluxes of cosmic neutrinos, undergoing flavor oscillations in such a gravitational background, which can be potentially measured by a terrestrial detector. Some astrophysical applications of our results are considered.
\end{abstract}

\section{Introduction}

Neutrinos are known to be massive and mixed particles. These neutrino properties lead to neutrino oscillations~\cite{Bil18}. There are various types of neutrino oscillations. We can mention neutrino flavor oscillations, when transitions between different flavor eigenstates happen, $\nu_\alpha \leftrightarrow\nu_\beta$, where $\alpha,\beta = e,\mu,\tau$. Transitions between different helicity states within one neutrino generation $\nu_-\leftrightarrow\nu_+$ are called neutrino spin oscillations.

Various external fields are established to influence the dynamics of neutrino oscillations. In this work, we are particularly interested how external gravitational fields can affect neutrino oscillations. Note that neutrino spin oscillations were previously studied in our work~\cite{Dvo06}. Here we shall study neutrino spin and flavor oscillations driven by a gravitational wave (GW). This interest is inspired by the direct observation of GWs reported in~\cite{Abb16}.

In this work, we summarize our recent studies in~\cite{Dvo19a,Dvo19b} of neutrino oscillations in GWs. First, in section~\ref{sec:SPIN}, we discuss neutrino spin oscillations in background matter, an external electromagnetic field, and GW. Then, in section~\ref{sec:FLAVOR}, we consider neutrino flavor oscillations in stochastic GWs emitted by randomly distributed sources. Some astrophysical applications are discussed.

\section{Neutrino spin oscillations in GW\label{sec:SPIN}}

In this section, we study the spin evolution of a massive neutrino in external fields in curved spacetime, neglecting the mixing between different neutrino types. Then, we take a particular gravitational background as GW, as well as background matter and a magnetic field. We derive the effective Schr\"odinger equation for neutrino spin oscillations and solve it numerically.

We consider one neutrino eigenstate, which is supposed to be a Dirac
particle, and neglect the mixing between different neutrino types.
The wave equation for a massive Dirac neutrino with the anomalous
magnetic moment, interacting with background matter and the electromagnetic
field $F_{\mu\nu}$ in curved spacetime, reads
\begin{equation}\label{eq:Direq}
  \left[
    \mathrm{i}\gamma^{\mu}\nabla_{\mu}-\frac{\mu}{2}F_{\mu\nu}\sigma^{\mu\nu}- 
    \frac{V^{\mu}}{2}\gamma_{\mu}
    \left(
      1-\gamma^{5}
    \right) -
    m
  \right]
  \psi = 0,
\end{equation}
where $\gamma^{\mu}=\gamma^{\mu}(x)$, $\sigma_{\mu\nu}=\tfrac{\mathrm{i}}{2}\left[\gamma_{\mu},\gamma_{\nu}\right]_{-}$,
and $\gamma^{5}=-\tfrac{\mathrm{i}}{4!}E^{\mu\nu\alpha\beta}\gamma_{\mu}\gamma_{\nu}\gamma_{\alpha}\gamma_{\beta}$
are the coordinate dependent Dirac matrices, $E^{\mu\nu\alpha\beta}=\varepsilon^{\mu\nu\alpha\beta}/\sqrt{-g}$
is the covariant antisymmetric tensor in curved spacetime, $g=\text{det}(g_{\mu\nu})$,
$g_{\mu\nu}$ is the metric tensor, $\nabla_{\mu}=\partial_{\mu}+\Gamma_{\mu}$
is the covariant derivative, $\Gamma_{\mu}$ is the spin connection,
$\mu$ is the magnetic moment of a neutrino, and $m$ is the neutrino
mass. The effective potential of the neutrino interaction with background matter $V^\mu$ can be found in the explicit form in~\cite{DvoStu02}. For example, in electroneutral hydrogen plasma, $V^0$ depends on the electron number density $n_e$.

We choose a locally Minkowskian frame, $x^{\mu}\to\bar{x}^{a}$ and $\eta_{ab}=e_{a}^{\ \mu}e_{b}^{\ \nu}g_{\mu\nu}$, where $e_{a}^{\ \mu}=\partial x^{\mu}/\partial\bar{x}^{a}$ are the vierbein vectors and $\eta_{ab}=\text{diag}(+1,-1,-1,-1)$. We also assume that the following identity is fulfilled: $\gamma_{abc}\eta^{bc}=e_{a\ ;\mu}^{\ \mu}=0$, where $\gamma_{cba}=e_{c\mu;\nu}e_{b}^{\ \mu}e_{a}^{\ \nu}=-\gamma_{bca}$
are the Ricci rotation coefficients. Then, equation~\eqref{eq:Direq} takes the form,
\begin{equation}\label{eq:Direqvftr}
  \left[
    \mathrm{i}\bar{\gamma}^{a}\partial_{a}-
    \frac{\mu}{2}f_{ab}\bar{\sigma}^{ab}-\frac{v^{a}}{2}\bar{\gamma}_{a}+
    \frac{v_{5}^{a}}{2}\bar{\gamma}_{a}\bar{\gamma}^{5}-m
  \right]
  \psi=0,
\end{equation}
where $\bar{\gamma}^{a}=e_{\ \mu}^{a}\gamma^{\mu}$ are the constant Dirac
matrices, $f_{ab}=e_{a}^{\ \mu}e_{b}^{\ \nu}F_{\mu\nu}$ and $v^{a}=e_{\ \mu}^{a}V^{\mu}$
are the corresponding objects expressed in the locally Minkowskian frame, and $v_{5}^{a}=v^{a}+\varepsilon^{abcd}\gamma_{cbd}/2$ is the effective axial-vector field.

Basing on equation~\eqref{eq:Direqvftr} and using the results of~\cite{DvoStu02}, we get the evolution equation for the invariant three vector of the neutrino spin $\bm{\zeta}$ in the form,
\begin{equation}\label{eq:nuspinrot}
  \frac{\mathrm{d}\bm{\zeta}}{\mathrm{d}t}=
  \frac{2}{\gamma}[\bm{\zeta}\times\mathbf{G}],
\end{equation}
where
\begin{equation}\label{eq:vectG}
  \mathbf{G}=\frac{1}{2}
  \left[
    \mathbf{b}_{g}+\frac{(\mathbf{e}_{g}\times\mathbf{u})}{1+u^{0}}
  \right]+
  \frac{1}{2}
  \left[
    \mathbf{u}
    \left(
      v^{0}-\frac{(\mathbf{vu})}{1+u^{0}}
    \right)-
    \mathbf{v}
  \right]+
  \mu
  \left[
    u^{0}\mathbf{b}-\frac{\mathbf{u}(\mathbf{u}\mathbf{b})}{1+u^{0}}+
    (\mathbf{e}\times\mathbf{u})
  \right].
\end{equation}
Here we represent $G_{ab}=\left( \gamma_{abc}+\gamma_{cab}+\gamma_{bca} \right) u^{c}=(\mathbf{e}_{g},\mathbf{b}_{g})$,  $u^{a}=(u^{0},\mathbf{u})$,
$v^{a}=(v^{0},\mathbf{v})$, $f_{ab}=(\mathbf{e},\mathbf{b})$, $\gamma=U^{0}$,
and $U^{\mu} = (U^0,\mathbf{U})$ is the neutrino four velocity in the world coordinates.

We take a plane circularly polarized GW, propagating along the $z$ axis, as a background gravitational field.
Choosing the transverse-traceless gauge, we get that the metric has
the form~\cite{Buo07}, 
\begin{equation}\label{eq:metric}
  \mathrm{d}s^{2}=g_{\mu\nu}\mathrm{d}x^{\mu}\mathrm{d}x^{\nu}=
  \mathrm{d}t^{2}-
  \left(
    1-h\cos\phi
  \right)
  \mathrm{d}x^{2}-
  \left(
    1+h\cos\phi
  \right)
  \mathrm{d}y^{2}+2\mathrm{d}x\mathrm{d}y h \sin\phi-\mathrm{d}z^{2},
\end{equation}
where $h$ is the dimensionless amplitude
of GW, $\phi=\left(\omega t-kz\right)$
is the phase of the wave, $\omega$ is frequency of the wave, and
$k$ is the wave vector. In equation~(\ref{eq:metric}), we use Cartesian
world coordinates $x^{\mu}=(t,x,y,z)$.

We consider the situation when neutrinos are emitted by the same source
of GWs. Moreover we suppose that, besides GW, a neutrino interacts with nonmoving
and unpolarized matter, i.e. $V^{0}\neq0$ and $\mathbf{V}=0$. We also take that a constant uniform magnetic
field transverse to the neutrino motion is present in the world coordinates
$x^{\mu}$. For example, we suppose that $\mathbf{B}=(B,0,0)$.

We consider the effective two component neutrino wave function $\tilde{\nu}=\exp\left[\mathrm{i}\sigma_{3}\left(\dot{\phi}t+\pi\right)/4\right]\nu$,
where $\nu^{\mathrm{T}}=(\nu_{+},\nu_{-})$, $\nu_{\pm}$ are the components describing different neutrino polarizations, and $\dot{\phi}=\left(\omega-kU_{z}/U^{0}\right)$. Taking
into account that $h\ll1$, we get that $\tilde{\nu}$ obeys the equation
\begin{equation}\label{eq:effSchrodtilde}
  \mathrm{i}\frac{\mathrm{d}\tilde{\nu}}{\mathrm{d}t}=
  \tilde{H}_{\mathrm{eff}}\tilde{\nu},
  \quad
  \tilde{H}_{\mathrm{eff}}=
  \left(
    \begin{array}{cc}
      -V^{0}/2 & \mu B
      \left[
        1-he^{-\mathrm{i}\dot{\phi}t}/2
      \right]
      \\
      \mu B
      \left[
        1-he^{\mathrm{i}\dot{\phi}t}/2
      \right] & V^{0}/2
    \end{array}
  \right).
\end{equation}
In equation~\eqref{eq:effSchrodtilde}, we assume that neutrinos are ultrarelativistic,
i.e. $U_{z}=\beta U^{0}$, where $\beta \approx 1$ is the neutrino
velocity.

\begin{figure}
  \centering
  \subfigure[]
  {\label{1a}
  \includegraphics[scale=.45]{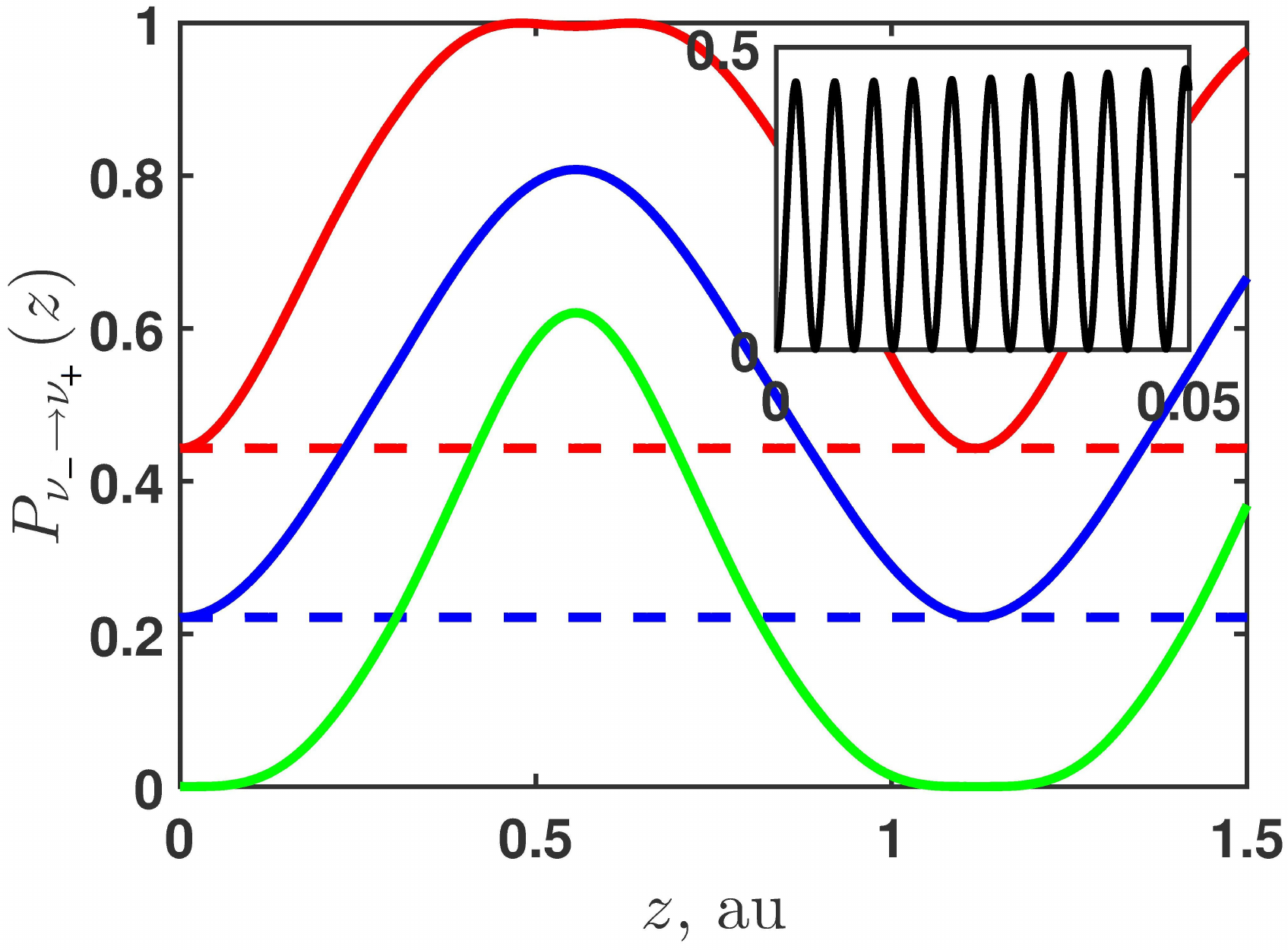}}
  \hskip-.2cm
  \subfigure[]
  {\label{1b}
  \includegraphics[scale=.45]{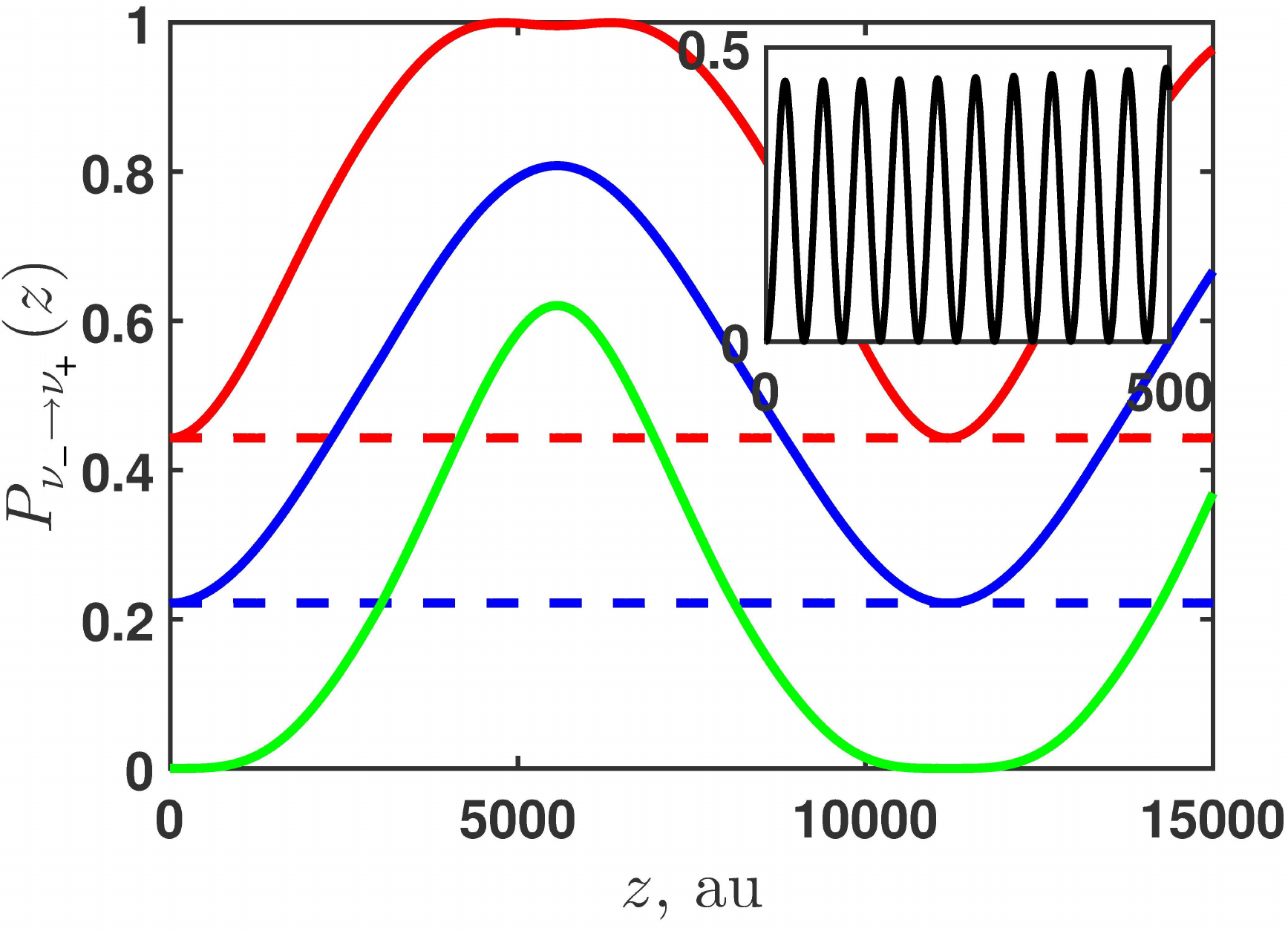}}
  \protect
  \caption{The probabilities $P_{\nu_{-}\to\nu_{+}}(z)$ for transitions
  $\nu_{-}\to\nu_{+}$ versus the distance $z=t$,
  passed by the neutrino beam, built on the basis of the numerical solution
  of equation~(\ref{eq:effSchrodtilde}) for different plasma densities, magnetic fields
  and neutrino energies.
  Red and green lines are the upper
  and lower envelope functions. Blue lines are the averaged transition
  probabilities. Solid lines correspond to $h=10^{-1}$ and dashed lines
  to $h=0$.
  (a) $n_e = 10^{22}\,\text{cm}^{-3}$, $\mu B = 5.8\times 10^{-16}\,\text{eV}$,
  and $E = 10\,\text{eV}$;
  (b) $n_e = 10^{18}\,\text{cm}^{-3}$, $\mu B = 5.8\times 10^{-20}\,\text{eV}$,
  and $E = 10^3\,\text{eV}$.
  The insets represent the transition probabilities, shown
  by the black lines, for $0<z<5\times10^{-2}\,\text{au}$ in panel~(a)
  and $0<z<5\times10^{2}\,\text{au}$ in panel~(b).
  \label{fig:PLR}}
\end{figure}

The Schr\"odinger equation analogous to equation~\eqref{eq:effSchrodtilde} was studied in~\cite{DvoStu04}. Using the results of~\cite{DvoStu04}, we suppose that $\dot{\phi}=2\Omega$, 
where $\Omega=\sqrt{(\mu B)^{2}+V_{0}^{2}/4}$ is the frequency of
the neutrino spin precession at the absence of GW, $\dot{\phi} = \omega m^2/2 E^2$ for relativistic neutrinos, and $E$ is the neutrino energy. The transition probability for $\nu_-\to\nu_+$ oscillations, based on the numerical solution of equation~\eqref{eq:effSchrodtilde}, is shown in figure~\ref{fig:PLR}. We have chosen the parameters of a neutrino and external fields corresponding to a particle propagating in the vicinity of merging black holes (BHs), surrounded by a dense magnetized accretion disk~\cite{Dvo19a}.

One can see in figure~\ref{fig:PLR} that the solid blue lines, which are the averaged transition probabilities, can reach a great values in contrast to dashed lines, which represent the corresponding transition probabilities at the absence of GW. It is the manifestation of the parametric resonance in neutrino spin oscillations.

\section{Neutrino flavor oscillations in GW\label{sec:FLAVOR}}

Now we study neutrino flavor oscillations under the influence of GW. We suppose that we deal with three flavor neutrinos $(\nu_e,\nu_\mu,\nu_\tau)$ which are related to the neutrino mass eigenstates $\psi_a$, $a=1,2,3$, with masses $m_a$, by means of the matrix transformation $\nu = U \psi$. These neutrinos are taken to interact with GWs.

The neutrino mass eigenstate was found in~\cite{For97}
to evolve in a gravitational field as
\begin{equation}\label{eq:psiaSa}
  \psi_{a}(\mathbf{x},t)\sim\exp
  \left[
    -\mathrm{i}S_{a}(\mathbf{x},t)
  \right],
\end{equation}
where $S_{a}(\mathbf{x},t)$ is the action for this particle, which
obeys the Hamilton-Jacobi equation,
\begin{equation}\label{eq:HJeq}
  g_{\mu\nu}\frac{\partial S_{a}}{\partial x_{\mu}}\frac{\partial S_{a}}{\partial x_{\nu}} =
  m_{a}^{2}.
\end{equation}
Here $g_{\mu\nu}$ is the metric tensor given in equation~\eqref{eq:metric}.

The solution to equation~\eqref{eq:psiaSa} in case of a plane GW was found in~\cite{Pop06}. Basing on the results of~\cite{Dvo19b,Pop06}, we get the contribution, linear in $h$, to the effective Hamiltonian for the neutrino mass eigenstates,
\begin{equation}\label{eq:Haag}
  (H_{m}^{(g)})_{aa}=-\frac{p^{2}h}{2E_{a}}\sin^{2}\vartheta\cos(2\varphi-\phi_{a}),
\end{equation}
where $E_{a}=\sqrt{m_{a}^{2}+p^{2}}$ is the neutrino energy, $\phi_{a}=\omega t(1-\beta_{a}\cos\vartheta)$
is the phase of GW accounting for the fact that a neutrino moves on
a certain trajectory, which is a straight line approximately, $\vartheta$ and $\varphi$ are the angles fixing the neutrino velocity with respect to the GW wave vector, and
$\beta_{a}=p/E_{a}$ is the neutrino velocity.

We have taken into account the contribution of GW linear in $h$ to
the diagonal elements of $H_{m}$. However, besides GW, there are
usual vacuum contributions to these elements, which have the form,
$(H_{m}^{(\mathrm{vac})})_{aa}=m_{a}^{2}/2E$. If we turn to the description of the evolution of the neutrino flavor
eigenstates, they obey the Schr\"odinger equation
$\mathrm{i}\dot{\nu}_{\lambda}=(H_{f})_{\lambda\kappa}\nu_{\kappa}$,
where the effective Hamiltonian takes the form, $H_{f}=UH_{m}U^{\dagger}$.

Let us consider the interaction of a neutrino with
a stochastic GW background. In this situation, following~\cite{LorBal94},
it is more convenient to deal with the density matrix $\rho$. We define $\rho_{\mathrm{I}}=U_{0}^{\dagger}\rho U_{0}$, where $U_{0}=\exp \left( -\mathrm{i}H_{0}t \right)$, $H_0$ is the time independent part of $H_{f}$, $H_{f} = H_0 + H_1(t)$, and $H_1(t)$ is the part of the Hamiltonian which incorporates the contribution of stochastic GWs.

We should average $\rho_{\mathrm{I}}$ over the directions of the
GW propagation and its amplitude. Then we consider the $\delta$-correlated
Gaussian distribution of $h$: $\left\langle h(t_{1})h(t_{2})\right\rangle =2\tau\delta(t_{1}-t_{2})\left\langle h^{2}\right\rangle $,
where $\tau$ is the correlation time. The evolution equation for
$\left\langle \rho_{\mathrm{I}}\right\rangle $, obtained in~\cite{Dvo19b},
has the form,
\begin{equation}\label{eq:rhoIeq3F}
  \frac{\mathrm{d}}{\mathrm{d}t}
  \left\langle
    \rho_{\mathrm{I}}
  \right\rangle = -
  \frac{3}{64}
  \langle h^2 \rangle
  \tau
  [ M, [M, \langle \rho_{\mathrm{I}} \rangle ] ],
\end{equation}
where
\begin{equation}\label{eq:Mdef3F}
  M = \frac{1}{2E} U \cdot \text{diag}
  \left(
    0, \Delta m_{21}^2, \Delta m_{31}^2
  \right)
  \cdot U^\dag.
\end{equation}
Here $\Delta m_{ab}^2 =  m_{a}^2 - m_{b}^2$ is the standard definition for the mass squared differences.

We use the results of~\cite{Ros11} to evaluate $\langle h^2 \rangle$ and $\tau$ in equation~\eqref{eq:rhoIeq3F}. If we take that stochastic GWs are emitted by coalescing supermassive BHs (SMBH) with masses up to $10^{10}M_\odot$, these parameters can be estimated as $\langle h^2 \rangle = 1.6\times 10^{-32}$ and $\tau \sim 10^{6}\,\text{s}$~\cite{Dvo19b}. 

\begin{figure}
  \centering
  \subfigure[]
  {\label{3a}
  \includegraphics[scale=.45]{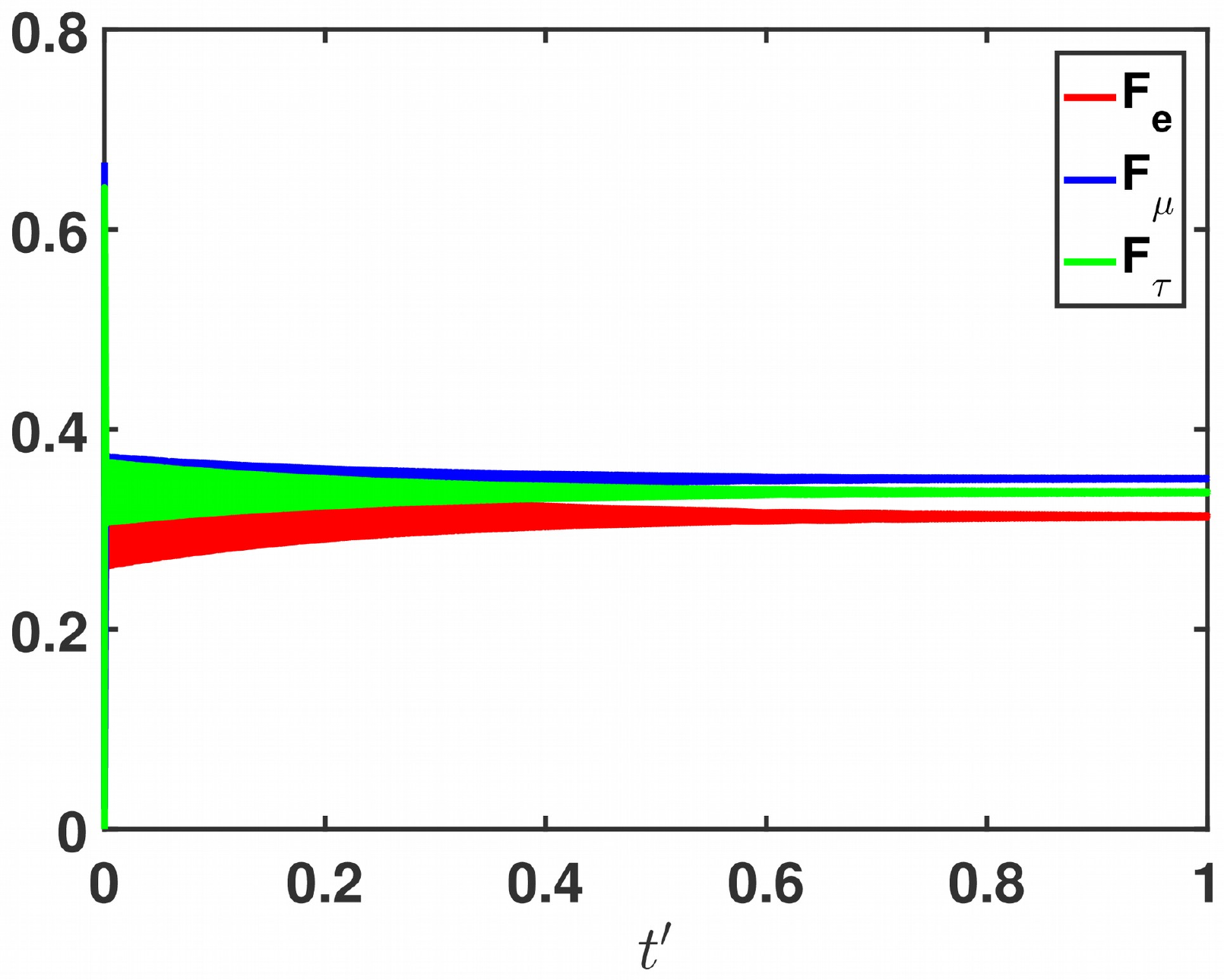}}
  \hskip0cm
  \subfigure[]
  {\label{3b}
  \includegraphics[scale=.45]{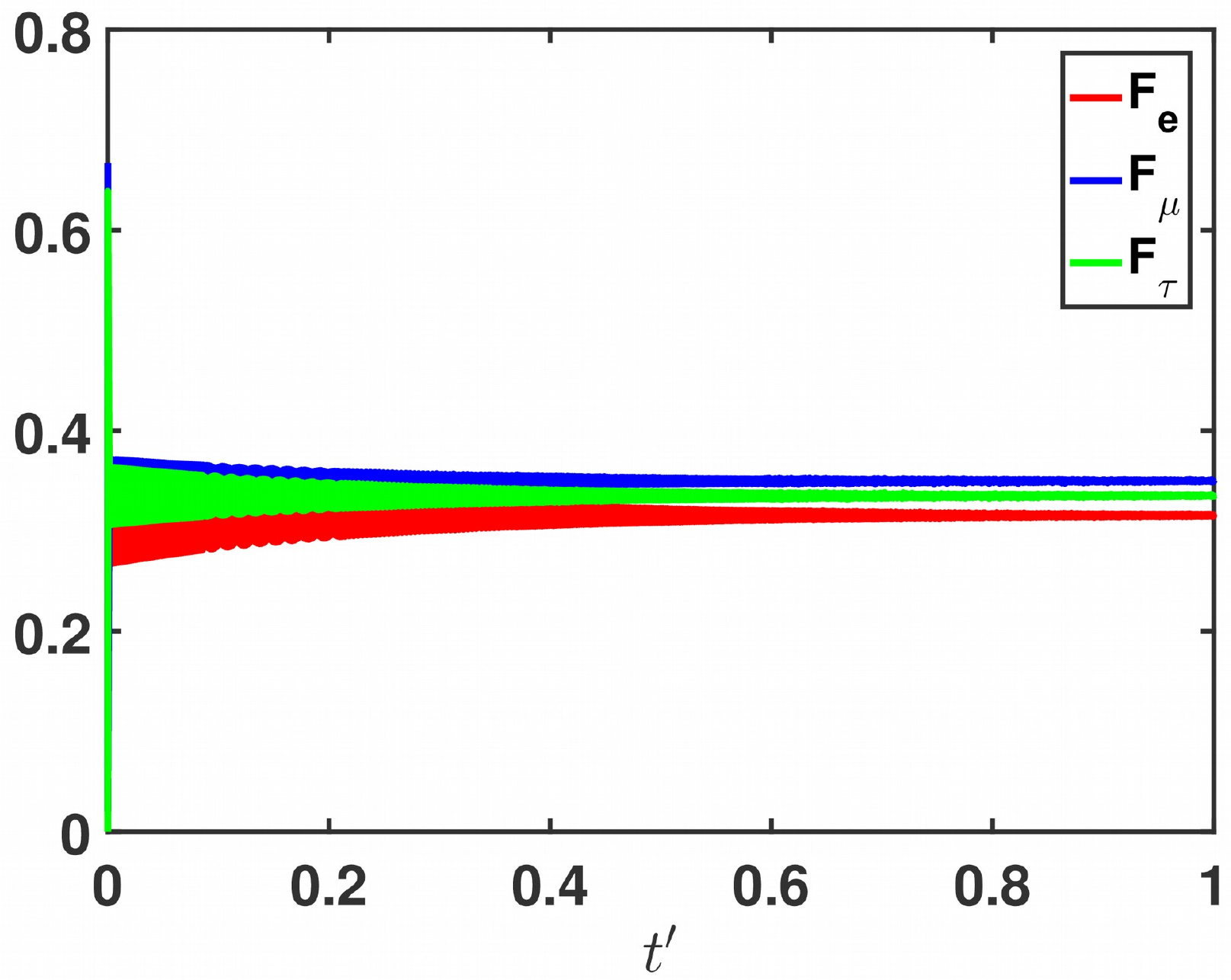}}
  \protect
  \caption{The  numerical solution
  of equations~\eqref{eq:rhoIeq3F} and~\eqref{eq:Mdef3F} 
  for three flavor neutrinos oscillations
  in stochastic GWs. The flux of electron neutrinos
  $F_e = \langle \rho_{11} \rangle$
  versus the dimensionless time $t' = t/L$   
  is shown by the red line, the flux of muon neutrinos $F_\mu = \langle \rho_{22} \rangle$
  is represented by the blue line, and the flux of tau neutrinos $F_\tau = \langle \rho_{33} \rangle$
  is depicted by the green line.
  The neutrino energy $E=0.5\,\text{MeV}$ and the propagation distance $L=1\,\text{Gpc}$.
  (a) Normal ordering with
  $\Delta m_{21}^2 = 7.39\times10^{-5}\,\text{eV}^2$,
  $\Delta m_{31}^2 = 2.53\times10^{-3}\,\text{eV}^2$,
  $\theta_{12} = 0.59$, $\theta_{23} = 0.87$, $\theta_{13} = 0.15$,
  and $\delta_\mathrm{CP} = 4.83$;
  (b) Inverted ordering with
  $\Delta m_{21}^2 = 7.39\times10^{-5}\,\text{eV}^2$,
  $\Delta m_{31}^2 = -2.51\times10^{-3}\,\text{eV}^2$,
  $\theta_{12} = 0.59$, $\theta_{23} = 0.87$, $\theta_{13} = 0.15$,
  and $\delta_\mathrm{CP} = 4.87$.
  \label{fig:fluxes3F}}
\end{figure}

In figure~\ref{fig:fluxes3F}, we show the solution of equations~\eqref{eq:rhoIeq3F} and~\eqref{eq:Mdef3F} for cosmic neutrinos with parameters, $\Delta m_{ab}^2$, the mixing angles $\theta_{ab}$, and the CP violating phase $\delta_\mathrm{CP}$, established in~\cite{Est19}. In figures~\ref{3a} and~\ref{3b}, we present the cases of both normal and inverted mass orderings. One can see that the fluxes reach the asymptotic values which do not coincide for different neutrino flavors.

The propagation length $L = 1\,\text{Gpc}$, taken in figure~\ref{fig:fluxes3F} for the fluxes to reach their asymptotic values, is comparable with the size of the visible universe.
Figure~\ref{fig:fluxes3F} is based on the initial condition (at a source) $(F_{e}:F_{\mu}:F_{\tau})_{\mathrm{S}}=(1:2:0)$. We can see in figure~\ref{3a} that, for the normal ordering, the asymptotic fluxes (at the Earth) are $F_{e\oplus} = 0.3127$, $F_{\mu\oplus} = 0.3504$, and $F_{\tau\oplus} = 0.3369$. For the inverted ordering, one has $F_{e\oplus} = 0.3154$, $F_{\mu\oplus} = 0.3497$, and $F_{\tau\oplus} = 0.3349$ in figure~\ref{3b}. It means that, at the Earth, the predicted fluxes are close to the case $(F_{e}:F_{\mu}:F_{\tau})_\oplus=(1:1:1)$. However, there is a small deviation from this prediction of~\cite{Bea03} for both normal and inverted mass orderings. Moreover, one can see that there is a small dependence of our results on the hierarchy of the neutrino masses.

The recent measurement of the flavor content of cosmic neutrinos was made in~\cite{Aar15}. 
Our prediction of the neutrino fluxes at a source in figure~\ref{fig:fluxes3F} is in the region not excluded in~\cite{Aar15}.
Of course,
neutrino energies in~\cite{Aar15}, $E>35\,\text{TeV}$, are
much higher than $E = 0.5\,\text{MeV}$ considered in our work (see figure~\ref{fig:fluxes3F}).

\section{Summary}

In this work, we summarize our recent achievements in~\cite{Dvo19a,Dvo19b} in the studies of neutrino oscillations in GWs. First, in section~\ref{sec:SPIN}, we have studied the evolution of the neutrino spin in background matter, electromagnetic and gravitational fields. Starting from the Dirac equation~\eqref{eq:Direq} for a massive neutrino in these external fields, we rederived the quasiclassical equations~\eqref{eq:nuspinrot} and~\eqref{eq:vectG} for the evolution of the neutrino spin, which was proposed previously in~\cite{Dvo13} basing on the equivalence principle. Then we have considered the neutrino motion in nonmoving and unpolarized matter, a transverse magnetic field, and a plane GW with the circular polarization. We have demonstrated that the parametric resonance in neutrino spin oscillations can take place in this system. Thus, the transition probability of neutrino spin oscillations can be significantly enhanced compared to the case when GW is absent. Some astrophysical applications have been discussed.

Then, in section~\ref{sec:FLAVOR}, we have studied neutrino flavor oscillations in GW. Using the results of~\cite{Pop06}, we have obtained the contribution of GW to the effective Hamiltonian for neutrino oscillations. Then, we have considered the neutrino interaction with stochastic GWs emitted by SMBHs. In this situation, we have solved the equation for the density matrix of flavor neutrinos. We have obtained that there is a small deviation from the fluxes ratio at a detector $(F_{e}:F_{\mu}:F_{\tau})_\oplus=(1:1:1)$ predicted in~\cite{Bea03}. The implication of our results for the observation of cosmic neutrinos has been discussed.

\ack

This work is performed under the government assignment for IZMIRAN. I am also thankful to RFBR (Grant No.~18-02-00149a) for a partial support.

\end{document}